\documentclass[aps,prb,showpacs, twocolumn, superscriptaddress]{revtex4}
\usepackage[utf8]{inputenc}
\usepackage[english]{babel}
\usepackage{graphicx}
\usepackage{color}
\usepackage{bm}
\usepackage{amssymb}
\usepackage[intlimits]{amsmath}
\usepackage{amsmath}

\newcommand{\be}{\begin{equation}}
\newcommand{\ee}{\end{equation}}
\newcommand{\ben}{\begin{equation*}}
\newcommand{\een}{\end{equation*}}

\begin{document}

\title{Impurity-induced magnetization in three dimensional antiferromagnet at quantum critical point}

\author{Y.~A.~Kharkov}
\email[E-mail: ]{y.kharkov@gmail.com}\affiliation{School of Physics, University of New South Wales, Sydney 2052,
Australia} 
\author{I.~S.~Terekhov}
\affiliation{School of Physics, University of New South Wales, Sydney 2052,
Australia}
\affiliation{Novosibirsk State University, Novosibirsk 630090, Russia}
\affiliation{Budker Institute of Nuclear Physics, Novosibirsk 630090,
Russia}
\author{O.~P.~Sushkov}
\affiliation{School of Physics, University of New South Wales, Sydney 2052,
Australia}

\begin{abstract}
We consider a single impurity with spin $S$ embedded in a three-dimensional antiferromagnetic system which is close to the quantum critical point (QCP), separating magnetically ordered and disordered phases. Approaching the QCP from the disordered phase we study the spatial distribution of spin density and staggered magnetization induced by the impurity. Using two methods (self-consistent Born approximation and renormalization group) we found a power law decay of the spin density  $\propto 1/r^3$, and of the staggered magnetization $\propto 1/r$ with relevant logarithmic corrections.
We demonstrate that the local spin at the impurity site $r=0$ approaches to zero at the QCP. 
We show that in the semiclassical limit of large $S$ the problem is equivalent to the exactly solvable independent boson model.
Our results demonstrate existence of spin-charge separation in the three dimensional systems in the vicinity of the QCP. 
\end{abstract}
\maketitle

\section{Introduction}

Quantum critical phenomena is extensively developing subject in modern condensed matter physics, in both theoretical and experimental frontiers.\cite{Sachdev:2011}  The most vivid manifestations of quantum phase transitions (QPT) arise in  low-dimensional systems such as cuprates and iron pnictides. However, quantum critical behaviour is also found in three-dimensional (3D+time) systems.
Well-known example of 3D compound with a magnetic quantum critical point (QCP) is TlCuCl$_3$.\cite{Tanaka:2003}  Under normal conditions the material is in the magnetically disordered phase, while pressure drives QPT to the antiferromagnetically ordered Neel phase. 

Quantum critical properties of a system can be significantly influenced by the presence of impurities. For instance, substitution of Cu atoms in the parent compound TlCuCl$_3$ with low concentration of nonmagnetic Mg impurities creates an uncompensated spin $1/2$ at the impurity site, which induces  magnetization around  the impurities and even lead to the formation of a long-range magnetic order in the macroscopic volume of the crystal.\cite{Tanaka:2000}
In the magnetically disordered phase the magnetization cloud around each impurity exponentially decays over a few lattice spacings from the impurity.
But  in the vicinity of the QCP the effect of impurity-induced magnetization can be notably enhanced.  Experimental observations reveal  an interplay between the impurity-induced staggered magnetization and a quantum criticality near the QCP. \cite{Imamura:2006, Suzuki:2009, Suzuki:2011}


Despite of the vast amount of theoretical work on the impurity-induced magnetization in quasi-1D  and 2D systems (see Refs. \cite{Sandvik:1997, Bobroff:2009, Roscilde:2010, Vojta:2000, Hoglund:2007} and references therein),  
we are not aware of similar studies in the relation to 3D materials. 
In the present paper we consider a single impurity with spin $S$ embedded in the 3D antiferromagnet (AF), which is close to the QCP, separating magnetically disordered and magnetically ordered phases. 
Conceptually, the problem is similar to the Kondo effect (see \cite{Kawakami, Affleck}), since, as we show below, the spin cloud screens the impurity's spin at the QCP (Bose-Kondo effect\cite{Florens}).
We study the spatial distribution of the nonlocal spin density and the staggered magnetization induced by the impurity. We show, that when approaching the QCP from the disordered phase,
the spin density around the impurity decays as $\propto 1/r^3$ with logarithmic corrections and the total spin accumulated in the delocalized cloud is equal to $S$.  
We also demonstrate that the induced staggered magnetization decays as $\propto 1/r$.

Closely related to the problem of  the impurity-induced
spin density and the impurity-induced Neel order is a
phenomenon of spin-charge separation (SCS). Conventional definition of SCS relies on the existence of two quasiparticles carrying spin and charge (”spinon” and ”chargon”), which is the case in 1D Tomonaga-Luttinger liquid of strongly interacting electrons.\cite{Tomonaga:1950, Luttinger:1963}.
By contrast, in higher spatial dimensions there are no known systems with SCS in the conventional definition. However, SCS exists in 2D
models, such as hole-doped AF \cite{Putikka:1994, Chen:1994, Tohyama:1996, Martins:1999}. Furthermore, the recent research \cite{Holt:2013} reports pronounced SCS in the vicinity of the magnetic QCP.
In the latter case, the precise meaning of SCS is different
from SCS in Tomonaga-Luttinger liquid. A hole creates
a spin cloud around the charge with radius which diverges at the QCP. As a result the
hole’s spin becomes delocalized and spatially separated from the impurity's charge pinned to the impurity's site, that basically means SCS.
In the present article we
show that such SCS also occurs in 3D systems near the QCP.

The paper is organized in the following way. In the Section II we introduce an effective field theory describing 3D AF doped with a single impurity in the vicinity of the QCP. Considering interaction of the doped AF with a probe magnetic field, we introduce an operator of the spin density and explain how we calculate the induced spin density.  Here we also provide method of calculation of the staggered magnetization around the impurity.
The rest of the paper is divided into two parts, which correspond to the two techniques of the calculations: Self-Consistent Born Approximation (SCBA) and Renormalization Group (RG) approach in $3+1$ dimensions. Section III refers to the calculation of the  spin density in SCBA for the most physically interesting case of the impurity with spin $S=1/2$. 
In the Section IV we again calculate the impurity-induced nonlocal spin density, the local spin of the impurity and the staggered magnetization using  RG technique. We also consider a semiclassical limit of the impurity with a large spin. We draw  our conclusions in the Section V.

\section{Effective Theory}\label{sec:Effective_Theory}

\begin{figure}[h]
\includegraphics[width=5cm]{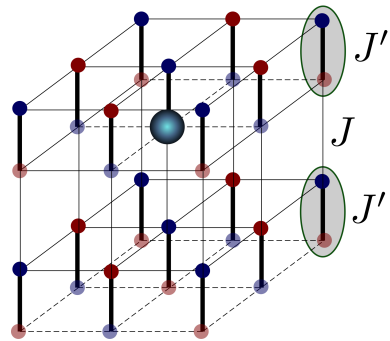}
\caption{An example of a lattice model for 3D AF with $O(3)$ QCP. Spins $S=1/2$ located at each site. Thin lines denote weak $J$ bonds and thick lines denote strong $J'$ bonds. A quantum phase transition between the Neel and the dimerized paramagnetic phases occurs at $(J'/J)_c=4.013$.\cite{Nohadani:2005, Jin:2012} The big blue sphere represents an impurity (hole) introduced into the lattice.}\label{fig:3D_AF_lattice}
\end{figure}

An example of a 3D lattice model which incorporates main features of magnetic quantum criticality  is presented in Fig. \ref{fig:3D_AF_lattice}.
The model corresponds to the cubic lattice  AF consisting of spins $S=1/2$ at each site with weak $J$ bonds and strong $J'$ bonds. The system has the QCP driven by parameter $g = J'/J$ and located at $g_c = 4.013$, which separates the disordered magnetic phase of spin dimers at $g>g_c$ from the Neel phase at  $g<g_c$.\cite{Nohadani:2005, Jin:2012}  This lattice model describes various properties of TlCuCl$_3$ near the pressure-driven QCP in zero and non-zero magnetic field.\cite{Nohadani:2005}

Substitution of a $S=1/2$ Cu$^{2+}$ ion with a spinless Mg$^{2+}$ creates a vacancy (hole). This is shown in Fig. \ref{fig:3D_AF_lattice}. 
The vacancy acts as an effective impurity with the spin $S=1/2$.
The vacancy in the lattice induces a nonlocal magnetization cloud around the impurity site. 
In the present paper we will calculate spatial distribution of the spin density and the staggered magnetization in the spin cloud around the impurity. 


Magnetic properties of the critical system are determined by low-energy magnetic excitations. The magnetic excitations are magnons in the Neel phase and triplons in the paramagnetic phase. Hereafter we use the term magnons for the both types of quasiparticles.
The effective theory, which describes magnons in the vicinity of the QCP is based on the following Lagrangian, see e.g. \cite{Zinn-Justin}:
\begin{eqnarray}\label{InitLagrangianMagn}
{\cal L}_M=\frac{\left(\partial_t\bm \phi\right)^2}{2}-\frac{\left(\nabla_i \phi_\mu\right)^2}{2}-\frac{\Delta_0^2\bm\phi^2}{2} - \frac{\alpha_0 (\bm \phi^2)^2}{4!} \,,
\end{eqnarray}
where $\phi_\mu = (\phi_x, \phi_y, \phi_z)$ is the magnon field, $\Delta_0^2 \propto g - g_c$ is the magnon gap (squared), $\alpha_0$ is a four-magnon coupling constant, $\partial_t$  is the time derivative,  $\nabla_i=\partial/\partial r_i$ is the three-dimensional gradient. 
Hereafter we set Plank constant and magnon speed equal to unity $\hbar = c = 1$. In the disordered magnetic phase  $\Delta_0^2>0$. Near the QCP the magnon gap $\Delta_0\rightarrow 0$.

The Lagrangian (\ref{InitLagrangianMagn}) contains quadratic terms as well as quartic term $\propto \phi^4$, describing the magnon self-action. 
The magnon self-action results in the
renormalization of the magnon gap $\Delta_0$ in the Lagrangian (\ref{InitLagrangianMagn}). From the one-loop RG calculations\cite{Zinn-Justin} it follows, that  in the disordered phase the evolution of the renormalized gap is given by 
\begin{equation}
\Delta^2 \propto \Delta_0^2 \left[\ln \frac{C(\Lambda)}{g-g_c}\right]^{ - \frac{N+2}{N+8}},
\end{equation}
where $N=3$ in the present case of the $O(3)$ universality class system, and $C(\Lambda)$ is a positive constant, determined by an ultraviolet scale $\Lambda$. 
Besides that, the $\phi^4$ term leads to renormalization of magnon quasiparticle residue.\cite{Zinn-Justin} 
However, the change of the residue appears only in the two-loop renormalization group, and therefore is small.   Hence, for our purposes we drop out the self-action term from the Lagrangian (\ref{InitLagrangianMagn}) and substitute the bare magnon gap to a renormalized value $\Delta_0\rightarrow \Delta$.

The Lagrangian of a non-interacting spin-$S$ impurity  reads 
\begin{eqnarray}\label{InitLagrangianElectron}
{\cal L}_{imp}&=&i\left(\psi^\dag (\bm r,t)\partial_t\psi(\bm r,t)-\left(\partial_t\psi^\dag (\bm r,t)\right)\psi(\bm r,t)\right)\,.
\end{eqnarray}
Here $\psi$ is the $2S+1$ component spinor. Hereafter we set the energy of the non-interacting impurity to zero. 
The Lagrangian which corresponds to the interaction between the impurity and  the magnon field in the disordered phase is \cite{Vojta:2000}  
\begin{eqnarray}\label{InitLagrangianInt}
{\cal L}_{int}=- \frac{\lambda}{S}\, \psi^\dag (\bm S\cdot\bm\phi)\psi\,,
\end{eqnarray}
where $\lambda$ is the coupling constant, $\bm S = (S_x, S_y, S_z)$ are the operators of the impurity spin acting in the $(2S+1)$-dimensional Hilbert space. 

The interaction of the impurity with magnons leads to appearance of the nonlocal part of spin density $\bm s(\bm r)$. In order to find $\bm s(\bm r)$ we use the Lagrangian of interaction of the system with external magnetic field (see, e.g.  \cite{Milstein:2011})
\begin{eqnarray}\label{eq:Lagrangian_B}
{\cal L}_B=-\left(\partial_t \bm \phi\cdot\left[\bm B\times \bm \phi\right] \right)+\frac{\left[\bm B\times \bm \phi\right]^2}{2} + \psi^\dag\left(\bm S \cdot\bm B\right)\psi\,.
\end{eqnarray}
We have set here $\mu_Bg=1$. Note, that Eq. (\ref{eq:Lagrangian_B}) is still valid, if the  magnetic field $\bm B$ is non-uniform. In the contrast to the majority of previous works, where $\bm B$ is considered uniform, the non-uniformity of the probe magnetic field is crucial for the present paper.
Linear in $\bm B$ terms in the Lagrangian (\ref{eq:Lagrangian_B}) provide following expression for spin density:
\begin{eqnarray}\label{eq:s_density_operator}
\bm s (\bm r)= \langle \frac{1}{2}\left([\bm \phi \times  \partial_t{\bm \phi} ] + \mathrm{h.c.} \right) + \psi^\dag \bm S \psi \rangle \,=\nonumber\\
\bm s_{nl}(\bm r) + \bm S_{imp} \delta(\bm r).
\end{eqnarray}
The brackets $\langle \cdots \rangle$ denote an averaging over the ground state of the system.
The term $ 1/2\langle[\bm \phi \times  \partial_t{\bm \phi} ] + \mathrm{h.c.} \rangle$ in Eq. (\ref{eq:s_density_operator}) is the  nonlocal part of the spin density $\bm s_{nl}(\bm r)$, induced by the impurity. The subscript "nl" stands hereafter for "nonlocal". The term $\langle\psi^\dag \bm S \psi \rangle$ in Eq. (\ref{eq:s_density_operator}) corresponds to the local spin $\bm S_{imp}$ at the impurity's site.  

In addition to the spin density we will consider the staggered magnetization, induced by the impurity. Writing down an Euler-Lagrange equation for the magnon field $\bm \phi$ from the action  $\int dt d^3 r\,\left\{\mathcal{L}_M+\mathcal{L}_{int}\right\}$  and taking expectation value of the result we obtain Yukawa-like form of the staggered  magnetization
\begin{equation}\label{eq:phi_def}
\langle \bm\phi(\bm r) \rangle = -\lambda \frac{e^{-\Delta r}}{4\pi r} \frac{\bm S_{imp}}{S}.
\end{equation}
At the QCP the exponent in Eq. (\ref{eq:phi_def}) is close to unity and  $\langle \phi(r) \rangle \propto 1/r$.
Therefore, in order to find corresponding prefactor, we only need to calculate the local spin  at the impurity site $\bm S_{imp}$.

To find local as well as  nonlocal components of the impurity-induced spin density we will proceed with the following procedure. 
We calculate the shift $\epsilon_B$ of the ground state energy, corresponding to the probe magnetic field $\bm B(\bm r') = \bm B \delta(\bm r'-\bm r)$. The energy shift of the system reads $\epsilon_B = \bm B \cdot \bm s (\bm r)$, therefore
\begin{equation}\label{eq:s(r)_deriv_B}
\bm s(\bm r) = \frac{\partial \epsilon_B}{\partial \bm B }\bigg\vert_{\bm B=0}.
\end{equation}

The spin density $\bm s(\bm r)=\bm e s(r)$ and the staggered magnetization $\langle\bm\phi(\bm r)\rangle = \bm e \langle \phi(r)\rangle$ are directed along the impurity spin $\bm S_{imp} = \bm e S_{imp}$ ($\bm e$ is a unit vector), 
and due to spatial isotropy of the system depend only on $r=|\bm r|$.
The ground state energy $\epsilon_g$ of the system is the position of a  singularity of the retarded impurity's Green's function $\hat G_B(\epsilon)$ and can be found from the Dyson's equation
\begin{eqnarray}\label{eq:groundEnergyInB}
\hat{G}^{-1}_B(\epsilon) = \epsilon-\hat\Sigma(\epsilon) - B^\mu \hat\Gamma^\mu(\epsilon, r) = 0\,,
\end{eqnarray}   
where $\hat \Sigma(\epsilon)$ is the self-energy of the impurity at zero magnetic field,
$\hat\Gamma^\mu(\epsilon, r)$ is the vertex function, corresponding to the interaction of the system with the probe  magnetic field. 
Note, that in Eq. (\ref{eq:groundEnergyInB}) we need to keep  only linear in $B^\mu$ terms.
From the rotational symmetry properties the only possible "kinematic" structure of the vertex
\begin{equation}
\hat\Gamma^\mu(\epsilon, r) = \Gamma(\epsilon, r) \hat S^\mu/S.
\end{equation} 

The vertex function can be split in local and nonlocal parts
\begin{equation}
\hat\Gamma^\mu(\epsilon, r) = 
\left\{
\begin{array}{l}
\hat\Gamma_{imp}^\mu(\epsilon), \quad r=0,\\
\hat\Gamma_{nl}^\mu(\epsilon, r), \quad r> 0.
\end{array}
\right. 
\end{equation}
\begin{figure}
\includegraphics[width=7cm]{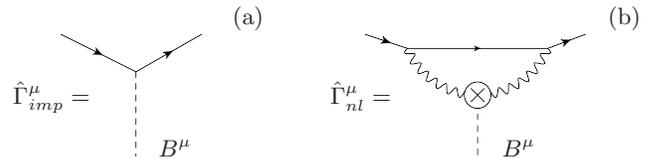}
\begin{picture}(0,0)
  \put(-137,50){\text{(a)}}  
  \put(-165,0){\text{$B^{\mu}$}}  
  \put(-220,20){\text{$\hat\Gamma_{imp}^\mu = $}}  
  \put(5,50){\text{(b)}}  
  \put(-35,0){\text{$B^{\mu}$}}  
  \put(-100,20){\text{$\hat\Gamma_{nl}^\mu = $}}  
\end{picture}
\caption{Example of diagrams for a coupling between a probe magnetic field $\bm B$ and (a) impurity spin, (b) nonlocal spin density. Solid line corresponds to a bare $(\lambda=0)$ impurity Green's function, waivy line represents magnon, dashed line represents probe magnetic field $\bm B$. The cross on the magnon line corresponds to the magnon-$\bm B$ vertex, provided by the term $\bm B[ \bm \phi\times\dot{\bm\phi}]$ in the Lagrangian (\ref{eq:Lagrangian_B}). }
\label{fig:Gamma}
\end{figure} 
Calculating the shift $\epsilon_B$  of the position of the singularity in the Green's function $\hat G_B(\epsilon)$ due to the probe magnetic field and using the formula (\ref{eq:s(r)_deriv_B}) we find the local and nonlocal components of spin density $s(r)$. 


Below we will calculate spin density using two approaches: Self-Consistent Born Approximation (for $S=1/2$) and Renormalization Group (for arbitrary $S$).

\section{Self-Consistent Born Approximation (S=1/2)} \label{sec:magnon_spin}

Standard approach for calculation of a single-fermion Green's function is $1/N$ expansion for  ${O}(N)$ group, where $N=3$ is the number of magnon components. Summation of leading terms in the  $1/N$  expansion results in Self-Consistent Born Approximation (SCBA), see Fig. \ref{fig:G_SCBA}. We will apply SCBA to the case of $S=1/2$ impurity only. As it will be demonstrated in Section \ref{sec:RG_S}, for $S>1/2$ corrections to impurity-magnon vertex, which are disregarded in SCBA, become relevant. Therefore, in this case SCBA fails and application of RG technique is necessary.

 \subsection{Impurity Green's function at zero magnetic field} 
\begin{figure}[h]
\includegraphics[width=8cm]{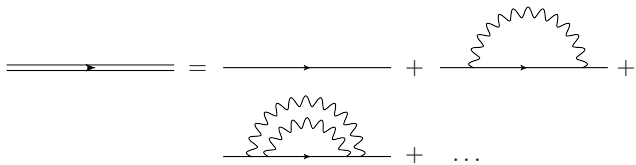}
\begin{picture}(0,0)
  \put(-162,33){\text{$=$}}
  \put(-80,33){\text{$+$}}
  \put(0,33){\text{$+$}}
  \put(-80,0){\text{$+ \quad \ldots$}}    
\end{picture}
\caption{Dyson's equation in SCBA. Double line is impurity's Green's function.}\label{fig:G_SCBA}
\end{figure}

To consider interaction of the system with the probe magnetic field, we first calculate 
Green's function of the impurity at zero magnetic field.
The Green's function of the impurity at $\bm B=0$ is proportional  to the identity matrix in the spin space $\hat G(\epsilon) = G(\epsilon)$. The Dyson's equation for the Green's function is graphically represented in Fig. \ref{fig:G_SCBA}. The analytical form of the
equation reads:
\begin{eqnarray}\label{eq:G_SCBA}
G(\epsilon)&=&\frac{1}{\epsilon-\Sigma(\epsilon)+i0}\, ,
\end{eqnarray} 
where the impurity self-energy is given by following expression
\begin{eqnarray}
\hat\Sigma(\epsilon)&=& \lambda^2 \int \frac{id\omega}{2\pi} \sum_{\bm q} \sigma^\mu \hat G(\epsilon-\omega) D_{\mu\nu}(\omega, \bm q) \sigma^{\nu}\nonumber =\\
&3&\sum_{\bm q}M^2_{\bm q} \hat G(\epsilon-\omega_{\bm q})\,.\label{eq:Sigma_zero_B}
\end{eqnarray}
Here $\omega_{\bm q}=\sqrt{\Delta^2 + \bm q^2}$ is the magnon dispersion, $M_{\bm q}=\lambda/\sqrt{2\omega_{\bm q}}$ is the matrix element corresponding to emission of magnon with momentum $\bm q$ by the impurity, $D_{\mu\nu}(\omega, \bm q)   = \delta_{\mu\nu}/(\omega^2 - \omega^2_{\bm q} +i0)$ is the magnon propagator. We expressed spin $1/2$ operators  via Pauli matrices $S^\mu = \sigma^\mu/2$. Combinatorial factor $3$ in Eq. (\ref{eq:Sigma_zero_B}) comes from summation over 
intermediate polarization state of magnon. 

The sum over momentum $\bm q$ in Eq. (\ref{eq:Sigma_zero_B}) diverges at large $|\bm q|$, therefore we have to  introduce ultraviolet cutoff $\Lambda$. The parameter $\Lambda$ depends on particular realisation of system and can be estimated as inverse lattice spacing in the host AF.


Solution to Dyson's equation (\ref{eq:G_SCBA}) near the QCP ($\Delta\rightarrow 0$) has the following form
\begin{equation}\label{eq:G_solution_SCBA}
G^{-1}(\epsilon) = (\epsilon - \epsilon_0 +i0)\sqrt{1+ \frac{3\lambda^2}{2\pi^2} \ln\left(\frac{\Lambda}{\epsilon_0 + \Delta - \epsilon - i0}\right)}
\end{equation}
in the vicinity of the singularity point $\epsilon_0 \approx -3\Lambda  \lambda^2/4\pi^2$. Formula (\ref{eq:G_solution_SCBA}) is obtained with logarithmic accuracy, i.e. assuming that $\ln \left (\frac{\Lambda}{\epsilon_0 + \Delta - \epsilon} \right) \gg 1$. 

The Green's function (\ref{eq:G_solution_SCBA}) has nontrivial analytic structure. At finite magnon gaps the quasiparticle pole at $\epsilon=\epsilon_0$ is separated by the gap $\Delta$ from the incoherent part of the Green's function. At the QCP, when $\Delta\rightarrow 0$ the pole and the branching point singularity are merging. The quasiparticle residue of the impurity Green's function $ G(\epsilon)$ vanishes in the vicinity of the QCP
\begin{equation}\label{eq:Z_SCBA}
Z = \left(1-\frac{\partial \Sigma(\epsilon_0)}{\partial\epsilon}\right)^{-1} = \frac{1}{\sqrt{1+\frac{3\lambda^2}{2\pi^2} \ln\left(\frac{\Lambda}{\Delta}\right)}}\Big|_{\Delta\rightarrow 0} \rightarrow 0.
\end{equation} 
Vanishing quasiparticle residue is the first signal of delocalization of the impurity-induced spin cloud and therefore indication of SCS. \cite{Holt:2013}

Typical value of the impurity-magnon coupling constant $\lambda$ can be estimated on the basis of the lattice model, shown in Fig. \ref{fig:3D_AF_lattice}. Lattice calculations\cite{Kharkov} result in the value of the effective coupling constant $\kappa = 3\lambda^2/2\pi^2 \sim 0.5$, appearing in front of the logarithm in formula (\ref{eq:G_solution_SCBA}). Therefore, the logarithmic corrections are significant in the vicinity of the QCP.

\begin{figure}[h]
\includegraphics[width=0.48\columnwidth]{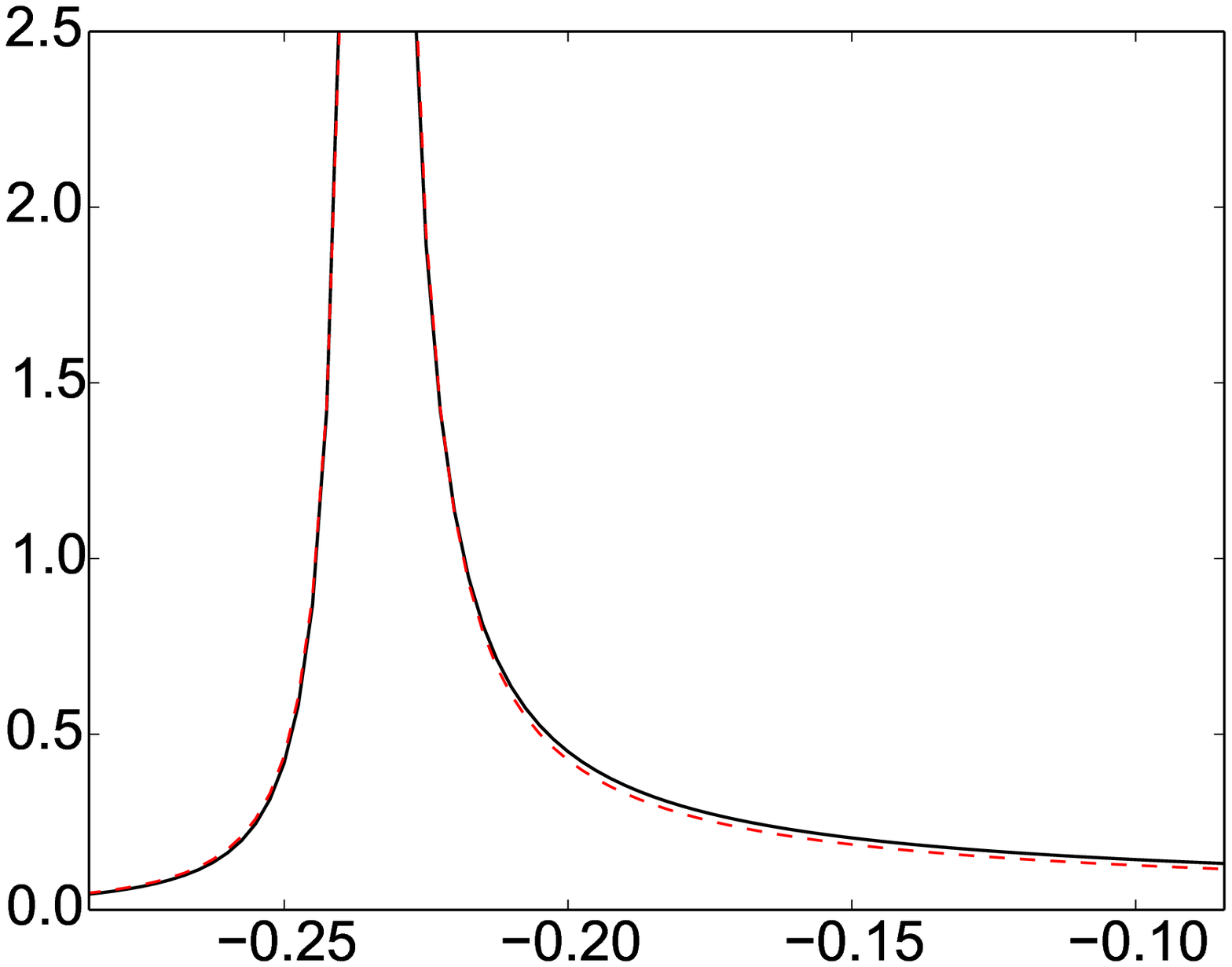}
\includegraphics[width=0.48\columnwidth]{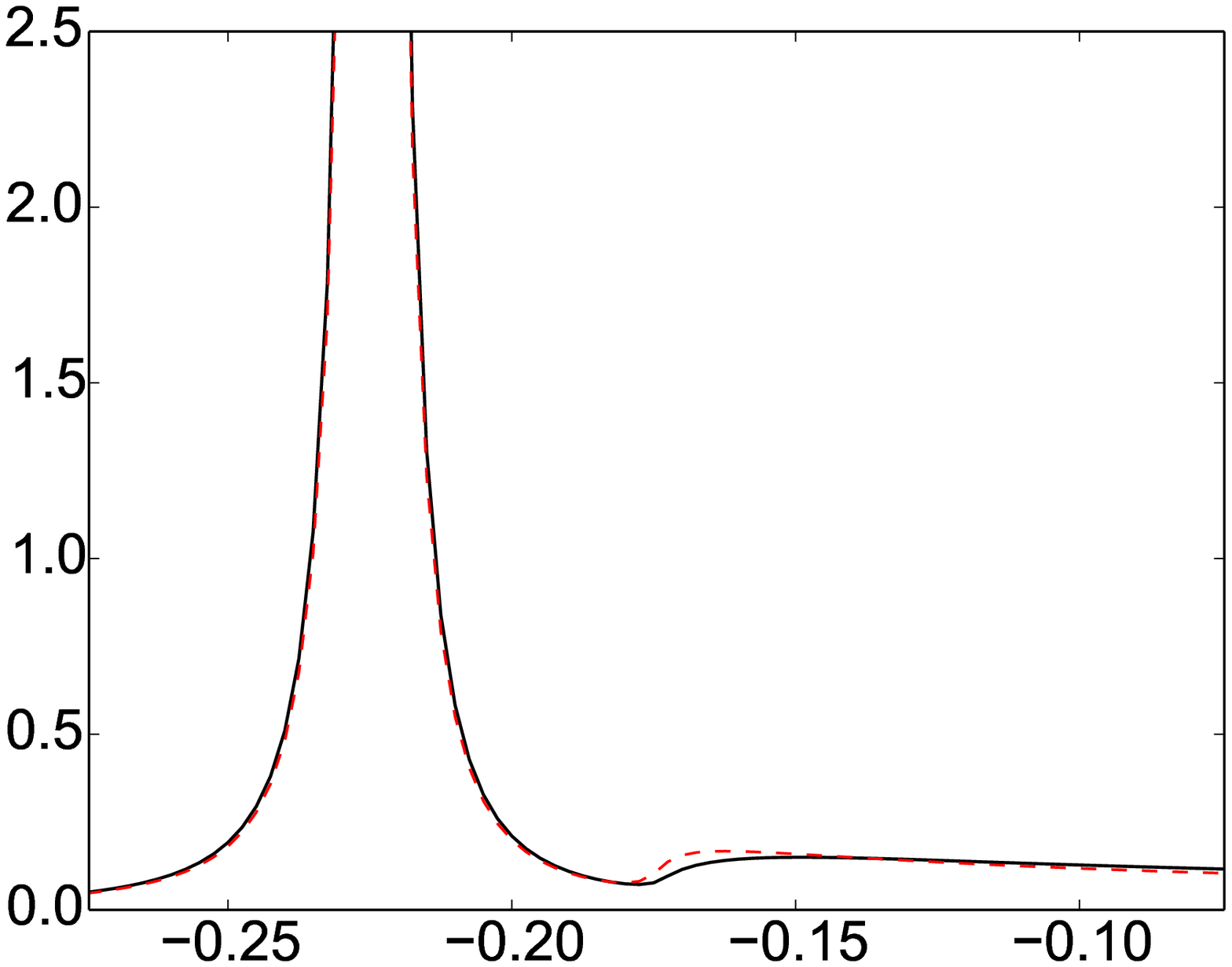}
\begin{picture}(0,0)
  \put(-150,70){\text{(a)}}
  \put(-30,70){\text{(b)}}
  \put(-65,-5){\text{$\epsilon/\Lambda$}}
  \put(-190,-5){\text{$\epsilon/\Lambda$}}  
  \put(-130,15){ \rotatebox{90}{\text{{\small$-1/\pi\, Im\, \{G(\epsilon)\}$}}}}
  \put(-250,15){ \rotatebox{90}{\text{{\small$-1/\pi\, Im\,\{G(\epsilon)\}$}}}}
\end{picture}
\caption{Spectral function of $S=1/2$ impurity obtained in SCBA.  Panel (a) corresponds to the QCP ($\Delta = 0$); panel (b) corresponds to magnon gap $\Delta = 0.05\Lambda$. Effective coupling constant is set to $\kappa = 0.6$. Solid black line corresponds to Green's function, calculated numerically, red dashed line corresponds to analytical formula (\ref{eq:G_solution_SCBA}). Note, that on the panel (a) the position of pole and branching point are merging.}\label{Fig:GreenFunction}
\end{figure}

Analytical result (\ref{eq:G_solution_SCBA}) for the impurity Green's function can be compared with direct numerical solution of the Dyson's equation (\ref{eq:G_SCBA}), corresponding plots for spectral functions $-1/\pi \, \textrm{Im}\{ G(\epsilon) \}$ are plotted in Fig. \ref{Fig:GreenFunction}.
An artificial broadening $i0\rightarrow i 2.5\times 10^{-3}\Lambda$ is introduced in the numerical procedure and in analytical formula (\ref{eq:G_solution_SCBA}).
We see an excellent agreement between the numerical and the analytical results.

Let us make a comment about the validity domain of SCBA for the results in the present section, and all following results, which will be derived in Sections \ref{sec:s_nl_SCBA}, \ref{sec:spin_impurity}. Formally, SCBA relies only on $1/N$ expansion of $O(N)$ group, independently on the value of the coupling constant $\lambda$. SCBA is applicable for arbitrary $\lambda$, in contrast to RG method, which works only for $\lambda<1$. We will  return to this discussion later, in the Section \ref{sec:RG_coupl_Z}. 

\subsection{Calculation of nonlocal  spin density $\bm s_{nl}(\bm r)$} \label{sec:s_nl_SCBA}
To evaluate nonlocal spin density induced by the impurity at the distances $r > 0$, we  
substitute the Green's function pole position $\epsilon_g = \epsilon_0 + \epsilon_B$  into Eq. (\ref{eq:groundEnergyInB}), expand it in $\epsilon_B$ up to the first order and use Eq. (\ref{eq:s(r)_deriv_B}). The result reads
\begin{eqnarray}\label{eq:s_nloc_ZG}
s_{nl}(r)=Z \Gamma_{nl}(\epsilon_0, r)\,.
\end{eqnarray}

Leading in coupling constant $\lambda$ contribution to 
the vertex $\hat\Gamma_{nl}^\mu$ is represented by the Feynman diagram shown in Fig. \ref{fig:Gamma}, (b). 
The analytical expression for the diagram is following
\begin{eqnarray}\label{eq:SigmaB_0}
&&\hat\Gamma^{\mu}_{nl}(\epsilon, r) = \Gamma_{nl}(\epsilon, r) \sigma^\mu = \int \frac{id\omega}{2\pi} \sum_{\bm q, \bm k}  \left(\lambda\sigma^\varkappa \right) \hat G_0(\epsilon - \omega)\nonumber\\ && \left(\lambda\sigma^\beta \right)\nonumber D_{\varkappa\nu}(\omega, \bm k) \left[-2i\omega \,  \varepsilon_{\mu \nu \alpha}\, e^{i(\bm q-\bm k)\bm r}\right]   D_{\beta \alpha}(\omega, \bm q)\, ,\nonumber\\
\end{eqnarray}    
where $\hat G_0(\epsilon) = 1/(\epsilon + i0)$ is the bare retarded Green's function of a non-interacting impurity. Expression in square brackets corresponds to magnon - probe magnetic field vertex, which we show in Fig. \ref{fig:Gamma}, (b) as a circle with a cross inside.

\begin{figure}[h]
\includegraphics[width=8cm]{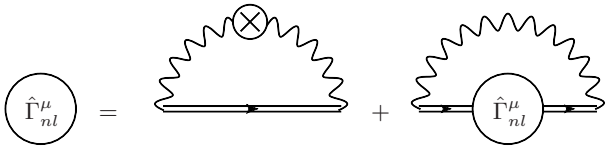}
\begin{picture}(0,0)
  \put(-223,10){\text{$\hat\Gamma_{nl}^\mu$}}  
  \put(-195,10){\text{$=$}}
  \put(-92,10){\text{$+$}}
  \put(-46,10){\text{$\hat\Gamma_{nl}^\mu$}} 
\end{picture}
\caption{Diagrammatic equation for "nonlocal" vertex function $\hat\Gamma_{nl}^\mu$.}\label{fig:Gamma_SCBA}
\end{figure}

SCBA equation for vertex $\hat\Gamma^\mu_{nl}(\epsilon, r)$ is graphically represented in Fig. \ref{fig:Gamma_SCBA}.
Analytical form of the equation for the vertex is
\begin{eqnarray}\label{eq:SigmaExact}
&&\Gamma_{nl}(\epsilon, r) = \Gamma_{nl}^{(0)} (\epsilon,  r) - \nonumber\\ &&\sum_{\bm q} M^2_{q} G^2(\epsilon - \omega_q) \Gamma_{nl}(\epsilon - \omega_q, r),
\end{eqnarray}
where $\Gamma_{nl}^{(0)}(\epsilon,   r)$ corresponds to the first term in the rhs of the diagrammatic equation in Fig. \ref{fig:Gamma_SCBA} and
reads
\begin{eqnarray}\label{eq:Gamma_0}
\Gamma_{nl}^{(0)} (\epsilon, r)=2\lambda^2 
 \sum_{\bm q, \bm k}    e^{i(\bm q-\bm k)\bm r} \frac{G(\epsilon - \omega_q) - G(\epsilon - \omega_k)}{\omega_q^2 - \omega_k^2} \,.
\end{eqnarray}
To obtain  expressions (\ref{eq:SigmaExact}) and (\ref{eq:Gamma_0}) we performed integration over $\omega$ in the rhs of the original SCBA equation, shown in Fig. \ref{fig:Gamma_SCBA}. Factor $(-1)$ in Eq. (\ref{eq:SigmaExact}) comes from algebraic identity for Pauli matrices $\sigma^\mu\sigma^\nu\sigma^\mu = - \sigma^\nu$.  Fformula (\ref{eq:Gamma_0}) follows from (\ref{eq:SigmaB_0}), where the bare impurity Green's function is changed to the "dressed" Green's function $G(\epsilon)$, shown in Fig. \ref{fig:G_SCBA}.

\begin{figure}
\includegraphics[width=9cm]{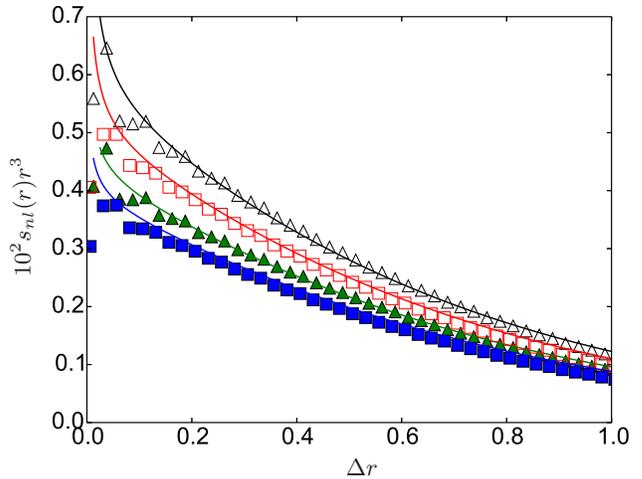}
\begin{picture}(0,0)
  \put(7,10){\text{$\Delta r$}}
  \put(-120,85){\rotatebox{90}{\text{$10^2 s_{nl}(r)r^3$}}}
\end{picture}
\caption{Spin density $s_{nl}(r)$ (multiplied by $r^3$) induced by $S=1/2$ impurity as a function of dimensionless distance $y=\Delta r$ calculated in SCBA. Points represent numerical results for different values of magnon gap and effective coupling constant $\kappa = 3\lambda^2/2\pi^2$. 
Squares correspond to $\Delta = 6.25 \times 10^{-3}\Lambda$, triangles correspond to $\Delta = 1.25\times 10^{-2} \Lambda$; filled markers represent $\kappa = 0.3$, open markers correspond to $\kappa = 0.6$.
Solid line is analytical approximation (\ref{eq:s_K}).}\label{FigSpinDistr}
\end{figure}

To find $s_{nl}(r)$ we solve numerically Eq. (\ref{eq:SigmaExact}) for the vertex $\Gamma_{nl}(\epsilon, r)$ and substitute the result together with the quasiparticle residue $Z$, obtained from numerical solution of Dyson's equation (\ref{eq:G_SCBA}), to Eq. (\ref{eq:s_nloc_ZG}). Solution to Eq. (\ref{eq:SigmaExact}) has been found iteratively, starting iterations from the  $\Gamma_{nl}(\epsilon, r)=\Gamma_{nl}^{(0)}(\epsilon, r)$. 
The results of the calculation of the spin density $s_{nl}(r)$ for different values of the  magnon gap $\Delta$ and coupling constant $\lambda$ are presented in Fig. \ref{FigSpinDistr}. 

For the purpose of computational efficiency  we used spherical cutoff $|\bm q|, |\bm k| \leq \Lambda $ in integrals in Eq. (\ref{eq:SigmaExact}) and (\ref{eq:Gamma_0}), instead of integrating over a cubic Brillouin zone. This cutoff scheme resulted in appearance of significant $r$-oscillations in the induced spin-density $s_{nl}(r)$, where the period of oscillations is $r\sim 1/\Lambda$ and the amplitude of the oscillations decaying with increasing $r$. It is clear that these oscillations are by-products of the rigid spherical cutoff and will be notably suppressed, if one performs proper 3D-integration over the cubic Brillouin zone. Hence, in Fig. (\ref{FigSpinDistr}) we plot numerical data for the spin-density, averaged over the period of the oscillations.

Our numerical calculations show that the starting approximation $\Gamma^{(0)}_{nl}(\epsilon,r)$ for vertex function and the solution $\Gamma_{nl}(\epsilon, r)$ of the SCBA equation (\ref{eq:SigmaExact}) are very close to each other. Therefore, to obtain analytical approximation for the nonlocal spin density we substitute (\ref{eq:Gamma_0}) in Eq. (\ref{eq:s_nloc_ZG}) and use formula (\ref{eq:G_solution_SCBA}) for the impurity's Green's function, the result reads
\begin{equation}\label{eq:s_K}
s_{nl}(r) = \frac{\lambda^2 \Delta }{4\pi^3\sqrt{1+\frac{3\lambda^2}{2\pi^2} \ln \frac{\Lambda}{\Delta} } \sqrt{1+\frac{3\lambda^2}{2\pi^2} \ln \Lambda r}} \frac{ K_1(2\Delta r)}{ r^2}.
\end{equation}
Here $K_1(x)$ is Macdonald function of the first kind. 
At distances $1/\Lambda< r < 1/\Delta$ using expansion of the Macdonald function $K_1(x)  \rightarrow 1/x$ at $x\rightarrow 0$,  we obtain power-law asymptotics  with logarithmic corrections for the spin density:
\begin{eqnarray}\label{eq:asympt_small_SCBA}
s_{nl}(r) \rightarrow  \frac{\lambda^2}{8\pi^3 r^3} \frac{1}{\sqrt{\left(1 + \frac{3\lambda^2}{2\pi^2} \ln \frac{\Lambda}{\Delta}\right)\left(1 + \frac{3\lambda^2}{2\pi^2} \ln {\Lambda r}\right)}}.
\end{eqnarray} 
At large distances $r>1/\Delta$ the spin density (\ref{eq:s_K}) is exponentially suppressed: $s_{nl} (r) \propto e^{-2 \Delta r}/r^{5/2}$.
In Fig. (\ref{FigSpinDistr}) solid lines correspond to the analytical result given by Eq. (\ref{eq:s_K}). One can see excellent agreement between the analytical and the numerical results.


%

\begin{figure}
\includegraphics[width=9cm]{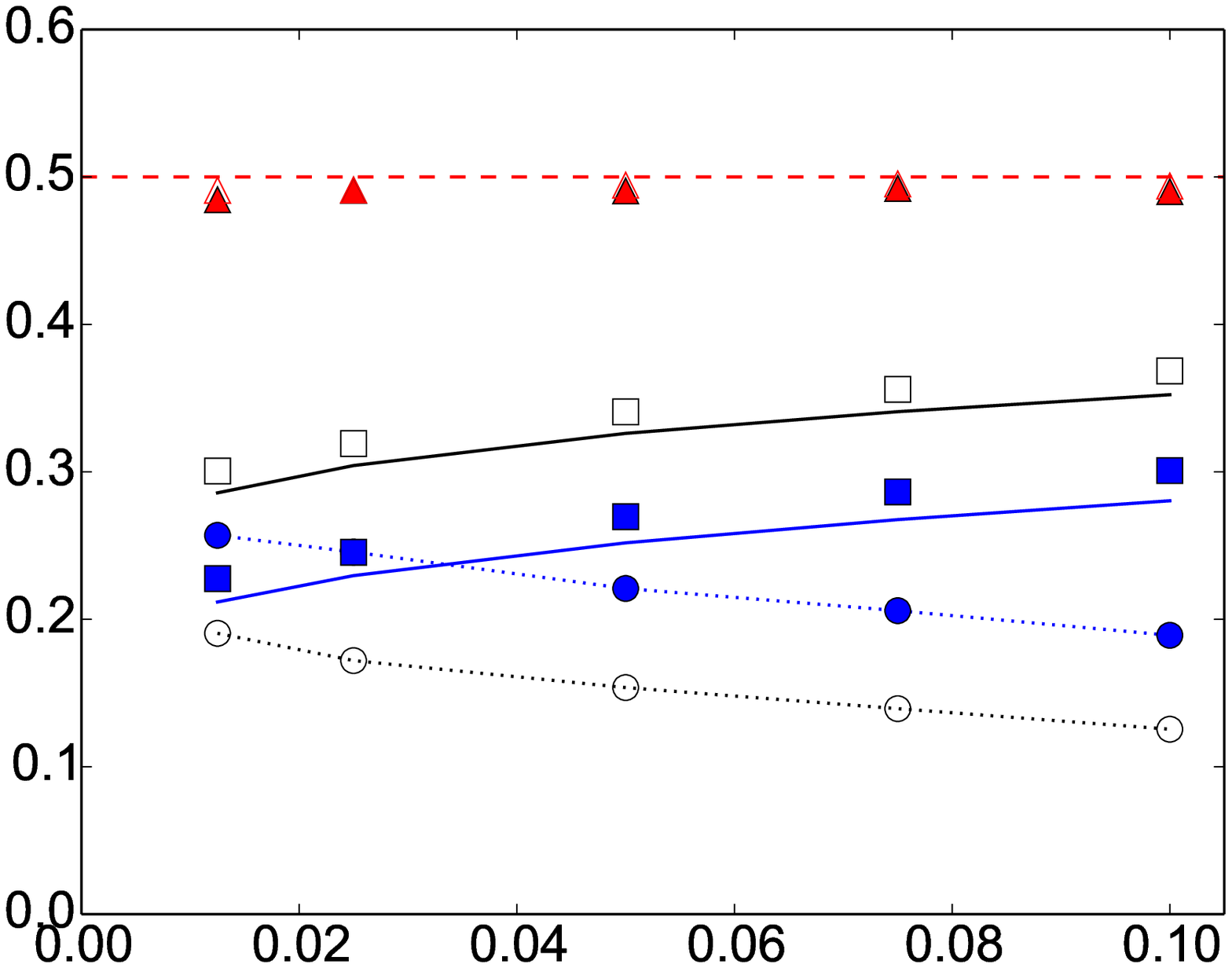}
\begin{picture}(0,0)
  \put(0,10){\text{$\Delta/\Lambda$}}
  \put(-120,85){\rotatebox{90}{\text{$S_{nl}$, $S_{imp}$}}}
\end{picture}
\caption{Integral nonlocal spin $S_{nl}$   and local spin of the imputity $S_{imp}$ as a function of the magnon gap $\Delta$ in SCBA.
Full and open markers correspond to values of effective coupling constant $\kappa = 0.6$ and $\kappa=0.3$. Circles correspond to $S_{nl}$,  squares represent $S_{imp}$ and triangles show the net spin $S_{nl}+S_{imp}$. Solid lines are theoretical predictions for the local impurity spin $S_{imp}$, given by Eq. (\ref{eq:S_imp_SCBA}). Dotted lines are visual guides for $S_{nl}$. Red dashed line corresponds to the net spin equal to $1/2$.}\label{FigSpinTot}
\end{figure}

The net spin of the system, which is given by the sum of local impurity spin and spin of nonlocal cloud is conserved and must be equal to $S=1/2$.
The integral spin, corresponding to the nonlocal spin density
\begin{eqnarray}\label{eq:S_nloc_def}
S_{nl}=\int d^3 r s_{nl}(r)\,
\end{eqnarray} 
is plotted in Fig. \ref{FigSpinTot} versus $\Delta/\Lambda$.
We use the numerical results for $s_{nl}(r)$, shown in Fig. \ref{FigSpinDistr}, in order to obtain $S_{nl}$.
One can see that the nonlocal spin $S_{nl}$ logarithmically increases with decreasing $\Delta$ and  tends to $S_{nl}=1/2$ at the critical point.  Therefore the rest of the spin should be attributed to the impurity spin $S_{imp}=1/2-S_{nl}$, which vanishes at the QCP. We check this statement in Section \ref{sec:spin_impurity}, calculating local spin of the impurity.

\subsection{Local spin of the impurity and staggered magnetization} \label{sec:spin_impurity}
\begin{figure}[h]
\includegraphics[width=6cm]{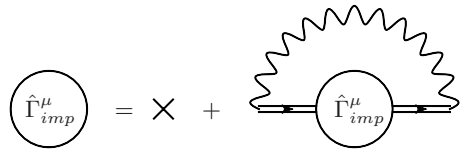}
\begin{picture}(0,0)
  \put(-168,12){\text{$\hat\Gamma_{imp}^\mu$}}  
  \put(-134,12){\text{$=$}}
  \put(-101,12){\text{$+$}}
  \put(-51,12){\text{$\hat\Gamma_{imp}^\mu$}} 
\end{picture}
\caption{Diagrammatic equation for "local" vertex function $\hat\Gamma_{imp}^\mu$. Cross represents bare vertex $\Gamma^{(0)}_{imp} = 1$.}\label{fig:Gamma_eps}
\end{figure}

To calculate impurity spin, localized at $\bm r=0$, we use similar approach to that we used in the previous section. We introduce local magnetic field $\bm B(\bm r) = {\bm B} \delta(\bm r)$ and calculate energy shift of the impurity due to the magnetic field.  
The result for the impurity spin reads
\begin{eqnarray}\label{eq:S_imp}
S_{imp} = \frac{Z}{2}\Gamma_{imp}(\epsilon)\Big\vert_{\epsilon = \epsilon_0}\,.
\end{eqnarray}     
Diagrammatic  equation for the  vertex function  $\hat\Gamma_{imp}^\mu(\epsilon) = \Gamma_{imp}(\epsilon)\sigma^\mu$ in SCBA  has the graphical representation, shown in Fig. \ref{fig:Gamma_eps}. 
Corresponding analytical form of the equation represented in Fig. \ref{fig:Gamma_eps} is
\begin{eqnarray}\label{eq:Gamma_eps}
\Gamma_{imp}(\epsilon) = 1 - \sum_{\bm q} M^2_{\bm q} G^2(\epsilon - \omega_{\bm q}) \Gamma_{imp}(\epsilon - \omega_{\bm q})\,.
\end{eqnarray}
Solving equation (\ref{eq:Gamma_eps}) and substituting the solution in to Eq. (\ref{eq:S_imp}) we obtain with logarithmic accuracy spin of the impurity
\begin{equation}\label{eq:S_imp_SCBA}
S_{imp} = \frac{1}{2\left(1 + \frac{3\lambda^2}{2\pi^2}\ln \frac{\Lambda}{\Delta}\right)^{2/3}}.
\end{equation} 
We also calculate residue spin of the impurity numerically, solving iteratively Eg. (\ref{eq:Gamma_eps}).
Both analytical and numerical results for $S_{imp}$ at different values of the parameters $\Delta$ and $\lambda$ are plotted in Fig. \ref{FigSpinTot}.
We see good agreement between the analytical and the numerical results.
From Fig. \ref{FigSpinTot} we can notice that the impurity spin logarithmically tends to zero when we approach to the QCP, $\Delta\to0$. 
In Fig. \ref{FigSpinTot} we also show the net spin $S_{imp}+S_{nl}$ for different values of $\Delta$ and $\lambda$. 

Results of our calculations, presented in Eq. (\ref{eq:asympt_small_SCBA}), Eq. (\ref{eq:S_imp_SCBA}) and also in Figs. \ref{FigSpinDistr}, \ref{FigSpinTot} show that at the QCP the local spin is approaching to zero and the spin of the system is accumulated in the nonlocal spin cloud. This delocalized spin cloud around impurity has size proportional to inverse magnon gap $r\simeq 1/\Delta$, and therefore significant part of impurity spin is separated from charge, localized at $r=0$. We will return to this discussion again in the Section \ref{sec:RG_s(r)}.

The net spin of the system equals to $1/2$. This is an exact statement and can be demonstrated at the diagrammatic level. One can trace mutual cancellations of corrections to the impurity spin and integral spin of nonlocal cloud in every order in $\lambda$. Corrections to the impurity spin $S_{imp}$ are cancelled by corrections to integral spin $S_{nl}$. The numerical results for the net spin of the system presented in Fig. \ref{FigSpinTot},  are consistent with the conservation of spin.

Using Eq. (\ref{eq:phi_def}) and Eq. (\ref{eq:S_imp_SCBA}) we obtain following expression for the staggered magnetization induced by the spin $1/2$ impurity
\begin{equation}\label{eq:phi_SCBA}
\langle \phi(r) \rangle = -\lambda\frac{e^{-\Delta r}}{4\pi r}\frac{1}{\left(1 + \frac{3\lambda^2}{2\pi^2}\ln \frac{\Lambda}{\Delta}\right)^{2/3}}.
\end{equation}
Away from the QCP  the staggered magnetization, induced by the impurity is exponentially small. In the vicinity of the QCP prefactor in Eq. (\ref{eq:phi_SCBA}) becomes logarithmically suppressed, however the staggered magnetization decays only as $\langle \phi(r) \rangle \propto 1/r$.

\section{Renormalization Group approach in 3+1 D}\label{sec:RG}
In this section we calculate nonlocal and local components of the spin density using RG technique in $3+1$ dimensions. In the RG approach the coupling constant $\lambda$ becomes dependent on the energy scale. 
Since $3+1$ D is the upper critical dimension, the evolution of the running coupling constant is logarithmic. It leads to logarithmic corrections to $s_{nl}$ and $\langle \phi(r) \rangle$, similar to results (\ref{eq:asympt_small_SCBA}) and (\ref{eq:phi_SCBA}) obtained in SCBA.  We derive our results for the case of an arbitrary spin  $S$ of the impurity in Sections \ref{sec:RG_coupl_Z} and \ref{sec:RG_s(r)}, and then we analyze the limit of a large spin $S$ in Section \ref{sec:RG_S}.

In RG technique we consider evolution of the coupling constant $\lambda$, quasiparticle residue $Z$, spin density and staggered magnetization with the energy scale $\mu$, starting evolution from the ultraviolet scale $\Lambda$ and finishing at the infrared  scale $\Delta$. The scale $\mu$ here has the meaning of the characteristic energy transfer from magnons to the impurity. At ultraviolet scale $\Lambda$ we set parameters of the theory to bare values, in our calculations $\Lambda$ plays a role of a renormalization point. Observables  in the vicinity of the QCP are calculated as the result of RG evolution from the ultraviolet scale  $\Lambda$ to the infrared scale $\mu = \Delta$.
\subsection{Evolution of coupling constant and quasiparticle residue}\label{sec:RG_coupl_Z}

\begin{figure}[h]
\includegraphics[width = 0.9\columnwidth]{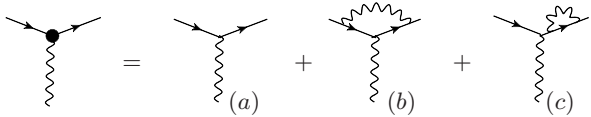}
\begin{picture}(0,0)
  \put(-180,15){\text{$=$}}
  \put(-115,15){\text{$+$}}
  \put(-55,15){\text{$+$}}

  \put(-140,0){\text{$(a)$}}
  \put(-80,0){\text{$(b)$}}
  \put(-20, 0){\text{$(c)$}}
  
\end{picture}
\caption{One-loop corrections to the impurity-magnon coupling constant $\lambda$.} 
\label{fig:Vertex_RG}
\end{figure}

First, we calculate evolution of the coupling constant $\lambda(\mu)$.  
One-loop correction to the coupling constant is represented by the sum of diagrams, shown in Fig. \ref{fig:Vertex_RG}.

Note, that in RG approach the correction to the coupling constant includes  the vertex correction [Fig. \ref{fig:Vertex_RG}, (b)], and also the self-energy correction [Fig. \ref{fig:Vertex_RG}, (c)]. 
This is different from SCBA, in which we disregard diagram (b).


Contribution $\delta\lambda^{(b)}$ to the coupling constant correction is given by the diagram (b) in Fig. \ref{fig:Vertex_RG} and reads
\begin{eqnarray}\label{eq:delta_coupl_1}
&&\bm S^\mu \delta\lambda^{(b)} =  \lambda^3  \frac{\bm S^\nu\bm S^\mu \bm S^\nu}{S^2} \times \nonumber \\ && \int \frac{id\omega'}{2\pi}\sum_{\bm k} G_0(\mu - \omega') G_0(\mu - \omega - \omega') D(\omega', \bm k)\approx \nonumber\\
&&\bm S^\mu \frac{\left(S(S+1)-1\right)}{S^2}\frac{\lambda^3}{4\pi^2} \ln \frac{\Lambda}{\mu}.
\end{eqnarray}
After cancelling out the factor $\bm S^\mu$ from the both sides of Eq. (\ref{eq:delta_coupl_1})  we obtain $\delta\lambda^{(b)}$.
The second contribution $\delta\lambda^{(c)}$, which comes from the  diagram (c) in Fig. \ref{fig:Vertex_RG} reads
\begin{eqnarray}\label{eq:delta_coupl_2_int}
\delta\lambda^{(c)} &&= \lambda^3 \left(1+\frac{1}{S}\right) G_0(\mu-\omega)\times \nonumber \\ &&\int \frac{id\omega'}{2\pi}\sum_{\bm k} G_0(\mu - \omega - \omega')D(\omega',\bm k)
\end{eqnarray}
and contains linear in $\Lambda$ and logarithmic in $\Lambda$ terms. The linear term corresponds to the shift of the position of quasiparticle pole $\epsilon_0$ in the impurity's Green's function and therefore is irrelevant for our purposes. The logarithmic term in $\delta\lambda^{(2)}$ reads 
\begin{equation}\label{eq:delta_coupl_2}
\delta\lambda^{(c)} \rightarrow -\frac{\lambda^3}{4\pi^2}\left(1+\frac{1}{S}\right)\ln\frac{\Lambda}{\mu}.
\end{equation}

Total correction to the coupling constant $\lambda$ is \begin{eqnarray}\label{eq:d_lambda_12}
\delta\lambda =  \delta\lambda^{(b)} + \delta\lambda^{(c)} = - \frac{\lambda^3}{4S^2\pi^2}\ln \frac{\Lambda}{\mu}.
\end{eqnarray}
Note, that for $S=1/2$ the vertex correction (\ref{eq:delta_coupl_1}) is suppressed by the factor $1/N = 1/3$, comparing to $\delta\lambda^{(c)}$. This suppression corresponds to standard $1/N$ expansion of $O(N)$ group.
However, at large $S$ the $1/N$ suppression of $\delta\lambda^{(b)}$ is
compensated by  $S$, and hence $\delta\lambda^{(b)}$ and $\delta\lambda^{(c)}$
to a large extent compensate each other,
$\delta\lambda^{(b)} \approx -\delta\lambda^{(c)}$.
Thus, at large $S$ the vertex correction becomes  significant and can not be disregarded. This is the reason why SCBA fails in the case of large impurity spin.
%
%

In the paradigm of RG, evolution of physical parameters on some energy scale $\mu$ is determined by the value of $\lambda(\mu)$ on the same scale. Hence, 
Eq. (\ref{eq:d_lambda_12}) results in the following  Gellman - Low equation
\begin{equation}\label{eq:d_ln_alpha}
\frac{d \lambda(\mu)}{d\ln\mu} = \frac{\lambda^3(\mu)}{4S^2\pi^2}.
\end{equation}
Solution to Eq. (\ref{eq:d_ln_alpha}) with the initial condition $\lambda(\Lambda) = \lambda$ is
\begin{equation}\label{eq:alpha_RG_solut}
\lambda(\mu) = \frac{\lambda}{\sqrt{1 + \frac{\lambda^2}{2S^2\pi^2} \ln\frac{\Lambda}{\mu}}}.
\end{equation}
Note, that the running coupling constant (\ref{eq:alpha_RG_solut}) vanishes in infrared limit:  $\lambda(\mu)\rightarrow 0$ at $\mu\simeq\Delta\rightarrow 0$. The RG scale $\mu$ is bounded from below by the value of magnon gap $\mu\geq\Delta$.

In order to find the quasiparticle residue of the impurity's Green's function we consider one-loop correction to the impurity's self-energy.  Logarithmic  part of this correction was already calculated  as a part of the diagram (c) in Fig. \ref{fig:Vertex_RG}.
Corresponding equation for evolution of $Z(\mu)$ reads
\begin{equation}\label{eq:d_ln_Z}
\frac{d \ln Z(\mu)}{d \ln\mu} =  \left(1+\frac{1}{S}\right) \frac{\lambda^2(\mu)}{4\pi^2}.
\end{equation}
Solution to Eq. (\ref{eq:d_ln_Z}) with initial condition $Z(\Lambda)=1$ reads
\begin{equation}\label{eq:Z_RG_solut}
Z(\mu) = \frac{1}{\left(1+\frac{\lambda^2}{2S^2\pi^2} \ln\frac{\Lambda}{\mu}\right)^{S(S+1)/2}} = \left(\frac{\lambda(\mu)}{\lambda}\right)^{S(S+1)}.
\end{equation}
The quasiparticle residue $Z(\mu)$ vanishes, while approaching to the QCP: $\mu\simeq\Delta\rightarrow0$.


Note, that RG approach, being used in the current Section, is valid if the effective coupling constant $\tilde\kappa=\lambda^2/2S^2\pi^2<1$, since we perform perturbative  expansion, such as in Eq. (\ref{eq:d_lambda_12}). However,  the proper expansion parameter in the vicinity of the QCP is not $\tilde\kappa$, but $\tilde\kappa \ln \Lambda/\Delta$.
RG method (in single-loop approximation) allows to sum up (leading) logarithmic corrections of the following kind $\tilde\kappa^m \sum_n \tilde\kappa^{n} \ln^n(\Lambda/\Delta)$. Therefore, the results obtained within one-loop RG in the Section \ref{sec:RG} are valid when $\tilde\kappa<1$ and $\tilde\kappa^2\ln ( \Lambda/\Delta)<1$, but the product $\tilde\kappa \ln (\Lambda/\Delta)$ can have an arbitrary value.
\subsection{Impurity spin and nonlocal spin density} \label{sec:RG_s(r)}
Now we consider RG evolution of the impurity spin $S_{imp}$ and spin density distribution $s(r)$ with renormalization scale $\mu$. 
As in Section \ref{sec:magnon_spin} we calculate $S_{imp}$ and $s(r)$, considering interaction of the system with probe magnetic field $\bm B(\bm r)$.

\begin{figure}[h]
\includegraphics[width = 0.9\columnwidth]{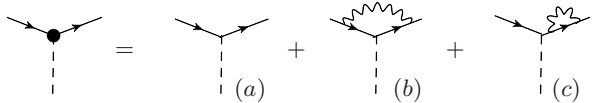}
  \put(-180,15){\text{$=$}}
  \put(-115,15){\text{$+$}}
  \put(-55,15){\text{$+$}}  
  \put(-135,0){\text{$(a)$}}
  \put(-75,0){\text{$(b)$}}
  \put(-15, 0){\text{$(c)$}}  
\caption{One-loop corrections to "local" spin $S_{imp}$. } 
\label{fig:Vertex_xi_RG}
\end{figure}

We start from calculation of corrections to  $S_{imp}$ due to interaction of the impurity with magnons. 
One-loop corrections to $S_{imp}$ are shown in Fig. \ref{fig:Vertex_xi_RG}.

Note, that the diagrams in Fig. \ref{fig:Vertex_xi_RG} are analogous to the diagrams in Fig. \ref{fig:Vertex_RG} for corrections to the coupling constant $\lambda$. The only difference is that the impurity-magnetic field coupling $\bm S_{imp}\cdot\bm B$ is proportional to the impurity spin $\bm S_{imp}$. Hence, RG evolution equation reads
\begin{equation}\label{eq:d_ln_xi}
\frac{d S_{imp}(\mu)}{d\ln\mu} = \frac{\lambda^2(\mu)}{4S^2\pi^2}S_{imp}(\mu).
\end{equation}
Solution to Eq. (\ref{eq:d_ln_xi}) with initial condition $S_{imp}(\Lambda)=S$ is following
\begin{equation}\label{eq:S_imp_RG}
S_{imp}(\mu) = \frac{S}{\sqrt{1+\frac{\lambda^2}{2S^2\pi^2} \ln \frac{\Lambda}{\mu}}}, 
\end{equation}
and proportional to solution (\ref{eq:alpha_RG_solut}) for running coupling $\lambda(\mu)$.
The local spin at the impurity site is equal to
$S_{imp}(\mu\simeq\Delta)$
and approaches to zero at the QCP. 
Using the result (\ref{eq:S_imp_RG}) and relation (\ref{eq:phi_def}) we obtain distribution of staggered magnetization around impurity
\begin{equation}\label{eq:phi_RG}
\langle \phi(r) \rangle =  -\frac{\lambda}{4\pi r} \frac{e^{-\Delta r}}{\sqrt{1+\frac{\lambda^2}{2S^2\pi^2} \ln \frac{\Lambda}{\Delta}}}.
\end{equation}

Now we calculate the nonlocal spin density $s_{nl}(r)$.
In RG technique it is more natural to use momentum representation for the spin density, therefore we write evolution equation for the Fourier component $s_{nl}(q)$. Leading in $\lambda^2$ contribution to $s_{nl}(q)$ is provided by one-loop diagram, shown in Fig. \ref{fig:Gamma}, (b). 
Evaluation of this diagram with logarithmic precision leads to
\begin{equation}\label{eq:s_0_RG}
s^{(0)}_{nl}(q) \approx \left\{ \begin{array}{l}
\frac{\lambda^2}{4S\pi^2} \ln \frac{\Lambda}{\mu}, \quad \mu\gg q, \\
\frac{\lambda^2}{4S\pi^2}\ln \frac{\Lambda}{q}, \quad \mu\ll q.
\end{array} \right. 
\end{equation}

Fourier transform of the second line of Eq. (\ref{eq:s_0_RG}) gives the spin density $s^{(0)}_{nl}(r) = \lambda^2/ 16S\pi^3 r^3$ at the distances $1/\Lambda < r < 1/\Delta$.  
In analogy with the result (\ref{eq:asympt_small_SCBA}), obtained in SCBA,  in RG calculations, we should expect logarithmic corrections to $\propto 1/r^3$ distribution. 
Note, that the logarithmic corrections are important, because they provide proper normalization condition of the integral nonlocal spin $\int d^3 r s_{nl}(r) \rightarrow S$  at the QCP. Volume integral of 
the spin density $\propto 1/r^3$ is logarithmically divergent $\propto \ln \Lambda/\Delta$ if we disregard the log corrections.

\begin{figure}[h]
\includegraphics[width = 0.95\columnwidth]{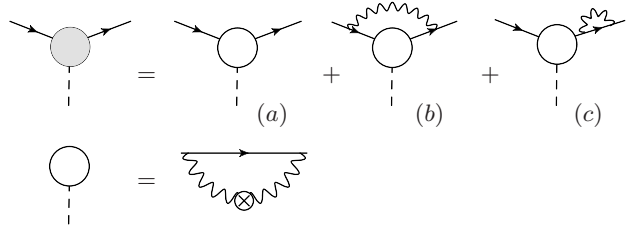}
  \put(-185,55){\text{$=$}}
  \put(-115,55){\text{$+$}}
  \put(-55,55){\text{$+$}}
  \put(-185,15){\text{$=$}} 
  \put(-140,40){\text{$(a)$}}
  \put(-80,40){\text{$(b)$}}
  \put(-20, 40){\text{$(c)$}} 
\caption{One-loop corrections to nonlocal spin density $s_{nl}(q)$.} 
\label{fig:Vertex_O_RG}
\end{figure}

In order to account for RG evolution of the spin-density, we evaluate single-loop corrections to the leading diagram presented in Fig.  \ref{fig:Gamma}, (b).  Diagrams (b) and (c) in Fig. \ref{fig:Vertex_O_RG} represent these corrections,  which are similar to corresponding diagrams in Figs. \ref{fig:Vertex_RG} and \ref{fig:Vertex_xi_RG}.
RG evolution of nonlocal spin density distribution reads
\begin{eqnarray}\label{eq:d_ln_s(q)}
\frac{d s_{nl}(q,\mu)}{d\ln\mu} = \left\{ 
\begin{array}{l}
\frac{\lambda^2(\mu)}{4S^2\pi^2} s_{nl}(q,\mu) - \frac{\lambda^2(\mu)}{4S\pi^2}, \quad \mu\gg q,  \\ 
\frac{\lambda^2(\mu)}{4S^2\pi^2} s_{nl}(q,\mu), \quad \mu\ll q.
\end{array}
\right.
\end{eqnarray} 

Note, that  equation for evolution of the spin density with $\mu$ is different in two domains $\mu\gg q$ and $\mu\ll q$, which is due to the fact, that one-loop expression  (\ref{eq:s_0_RG}) for spin density has various form in the both domains.
We solve Eq. (\ref{eq:d_ln_s(q)}) separately in the two domains and match the solutions at $\mu\simeq q$. As an initial condition for the evolution equation (\ref{eq:d_ln_s(q)}) we set $s_{nl}(\Lambda) = 0$. We obtain following result for the spin density
at infrared scale $\mu\simeq\Delta$:
\begin{equation}\label{eq:s(q)_RG}
s_{nl}(q) = \frac{S}{\sqrt{1+\frac{\lambda^2}{2S^2\pi^2} \ln \frac{\Lambda}{\Delta}}} \left( \sqrt{1+\frac{\lambda^2}{2S^2\pi^2} \ln \frac{\Lambda}{q}} - 1 \right).
\end{equation}
Condition of the net spin conservation in the momentum representation has the form $s_{nl}(q)\big\lvert_{q\rightarrow 0} + S_{imp}   = S$. Using expressions (\ref{eq:S_imp_RG}) and (\ref{eq:s(q)_RG}) it is easy to check the net spin conservation, having in mind that the low bound for the momentum $q$ in our formulas is $q\simeq\Delta$. 

Calculating Fourier transform of Eq.  (\ref{eq:s(q)_RG}) we obtain spatial distribution of the induced spin density 
\begin{equation}\label{eq:s(r)_RG}
s_{nl}(r) = \frac{\lambda^2}{16S\pi^3 r^3\sqrt{1+\frac{\lambda^2}{2S^2\pi^2} \ln \frac{\Lambda}{\Delta}} \sqrt{1+\frac{\lambda^2}{2S^2\pi^2} \ln \Lambda r}}
\end{equation}
at the distances $1/\Lambda < r< 1/\Delta$. 
Using (\ref{eq:s(r)_RG}) and (\ref{eq:S_imp_RG}) one can verify conservation of the net spin in $r$-representation: $\int d^3r\, s_{nl}(r)+S_{imp} = S$. Integration of nonlocal spin density should be performed in the range of the distances $1/\Lambda < r < 1 /\Delta$ which is defined by the infrared and ultraviolet cutoffs for our theory.

Note, that at the QCP the main contribution to the nonlocal spin $\int d^3r\, s_{nl}(r)$ comes from large distances $r<1/\Delta \rightarrow \infty$. 
Indeed, the integral
\begin{equation}\label{eq:RG_Integral_Spin}
\int_{1/\Lambda \leq r\leq R} d^3r\, s_{nl}(r) =  S\frac{\sqrt{1+\frac{\lambda^2}{2S^2\pi^2} \ln \Lambda R} - 1}{\sqrt{1+\frac{\lambda^2}{2S^2\pi^2} \ln \frac{\Lambda}{\Delta}}}
\end{equation}
logarithmically grows as a function of the upper integration limit $R$, which means that major part of spin in the nonlocal cloud is accumulated at the distances of the order of $R \simeq 1/\Delta$.
At the same time, the local spin of the impurity $S_{imp}$ vanishes at the QCP, see Eq. (\ref{eq:S_imp_RG}). Therefore, we conclude that at the QCP the impurity's spin is spatially separated from the impurity charge.

The results (\ref{eq:Z_RG_solut}), (\ref{eq:phi_RG}) and (\ref{eq:s(r)_RG}) obtained in RG technique are similar to corresponding answers (\ref{eq:Z_SCBA}), (\ref{eq:phi_SCBA}) and (\ref{eq:asympt_small_SCBA}), obtained in SCBA. For the spin $S=1/2$ the difference is in the numerical factors in front of the logarithms: $3\lambda^2/2\pi^2$ in SCBA, comparing to  $2\lambda^2/\pi^2$ in RG.
For $Z$ and $\langle \phi(r)\rangle$ the powers of logarithms are also insignificantly changed: $1/2\rightarrow 3/8$ and $2/3\rightarrow 1/2$, respectively.
The reason for this minor changes  is due to the $1/N$ vertex correction, which is accounted in RG approach (see diagram (b) in Fig. \ref{fig:Vertex_RG}), and is disregarded in SCBA. 
The RG results are more accurate, then SCBA results.
However, the expansion of the RG results and  SCBA results coincide up to the single-loop order (first order in $\lambda^2$).


\subsection{Semiclassical limit: impurity with large spin} \label{sec:RG_S}

From theoretical point of view it is interesting to consider the semiclassical case of large spin of the impurity. Taking formal limit $S\rightarrow\infty$ in Eqs. (\ref{eq:S_imp_RG}),  (\ref{eq:phi_RG}) and (\ref{eq:s(r)_RG}) we obtain
\begin{eqnarray}\label{eq:large_spin}
S_{imp} = S, \quad s_{nl}(r) = 0, \quad \langle\phi(r)\rangle = -\lambda \frac{ e^{-\Delta r}}{4\pi r}.
\end{eqnarray}
We see from Eq. (\ref{eq:large_spin}), that in the semiclassical limit there is no nonlocal spin density around impurity and local spin $S$ is "unscreened" in this case. Therefore, there is no spin-charge separation in the semiclassical limit.

Note, that local impurity spin, nonlocal spin density and staggered magnetization in the semiclassical limit are provided just by tree-level approximation.  The reason is that quantum fluctuations of the of impurity spin are suppressed at large $S$. 
Indeed, let us consider the case of impurity in the state with maximal projection of spin on the quantization axis $z$: $|S,S_z=S\rangle$. Interaction of the impurity with magnon either leaves projection $S_z$ to be unchanged or changes it by unity, $\Delta S_z=-1$.
The action of operator $\hat S_z$ on the state $|S,S\rangle$  provides eigenvalue $S$. On the other hand, matrix element of the lowering operator $\hat S_{-}$ between states $|S,S-1\rangle$ and $|S,S\rangle$ is equal to $\sqrt{2S}$. Therefore, the processes with change of projection of the impurity spin are suppressed in the limit of large $S$.

In the semiclassical limit only $z$-component of the operator of impurity spin  is relevant, as a result the Lagrangian in Eq. (\ref{InitLagrangianInt}), corresponding to interaction of the impurity with magnons takes the form
\begin{equation}\label{eq:L_int_class}
\mathcal{L}_{int} = -\lambda \psi^\dag\psi \phi_z.
\end{equation}
Thus, the problem of classical impurity "dressed" with $z$-polarized magnons is equivalent to the problem of impurity interacting with scalar bosonic field $\phi_z$. 



The problem of interaction between impurity and scalar boson field is known as independent boson model, this model is exactly solvable.\cite{Mahan}
The exact solution agrees with Eq. (\ref{eq:large_spin}).


Retarded Green's function of the impurity in time representation at $t>0$  reads \cite{Mahan}
\begin{equation}\label{eq:G_Mahan}
G(t) = -i\exp\left[it\epsilon_0 - \lambda^2\sum_{\bm q}\frac{(1-e^{-i\omega_q t})}{2\omega_q^3}\right],
\end{equation}
where $\epsilon_0 = -\lambda^2 \sum_q 1/2\omega_q^2$.
Performing Fourier transformation of the impurity Green's function (\ref{eq:G_Mahan}), and calculating quasiparticle residue   at the  Green's function pole $\epsilon = \epsilon_0$, we obtain
\begin{equation}\label{eq:large_spin_Z}
Z = \exp\left(-\lambda^2\sum_q\frac{1}{2\omega_q^3}\right)=\left(\frac{\Delta}{\Lambda}\right)^{\lambda^2/4\pi^2}.
\end{equation}
In the limit $S \to \infty$ the RG result (\ref{eq:Z_RG_solut}) is consistent with Eq. (\ref{eq:large_spin_Z}).


%



%

\section{Discussion of results and conclusion} \label{sec:conclusion}

In the present paper we have considered a single impurity
with spin S embedded into 3D AF system, which is close
to the $O(3)$ quantum critical point (QCP), separating paramagnetic 
and Neel magnetic phases. The impurity spin induces the usual magnetization
and the staggered magnetization clouds around position of the impurity.
Using the effective Lagrangian method and approaching the QCP from the 
disordered  phase, we have calculated spatial distributions
of  the spin density $s(r)$ (magnetization) and the staggered 
magnetization $\langle \phi(r) \rangle$ in the cloud.
For calculations we use two different methods, Self Consistent Born
Approximation (SCBA) and Renormalization Group (RG).
SCBA is justified by the small parameter $1/N$ where $N=3$ for the $O(3)$ group,
while RG is justified by the small coupling constant.
We show that for $S=1/2$ results of both methods are consistent within
expected accuracy $1/N$. However, at larger values of the impurity spin
the SCBA method is not valid because the small parameter $1/N$ is
compensated by the large spin. Therefore, for $S\geq 1$ only RG results
are valid.

The impurity quasiparticle residue vanishes at the QCP, see Eq. (\ref{eq:Z_RG_solut}). 
This is the first indication, that the impurity spin is fully transferred to 
the magnon cloud. 
The effect of screening of impurity's spin by spin-one magnetic fluctuations  is a Kondo effect in a bosonic sector. \cite{Florens}
The spin density has a local component $S_{imp}\delta(\bm r)$, which is
localized at the site of the impurity, as well
as a spatially distributed nonlocal part $s_{nl}(r)$. 
Because of the vanishing residue the average impurity spin $S_{imp}$ 
logarithmically vanishes at the QCP, see Eq. (\ref{eq:Z_RG_solut}).
Of course, the total spin $S$ is conserved and it is transferred into
the nonlocal spin cloud. The nonlocal spin  density at $r < 1/\Delta$,  
where $\Delta$ is the magnon gap, decays as $s_{nl}(r)\propto 1/r^3$ 
with proper logarithmic corrections, see Eq. (\ref{eq:s(r)_RG}).
Obviously, at $r > 1/\Delta$ the spin density  decays exponentially.

Integral spin in the nonlocal spin cloud is mainly accumulated at large distances $r \simeq 1/\Delta$, see Eq. (\ref{eq:RG_Integral_Spin}).
Therefore the spin is spatially separated from the impurity
and at $\Delta \to 0$ the separation scale becomes infinite.
In this sense our results demonstrate the spin-charge separation in 3D
magnetic systems at the QCP.

Interestingly, the cloud of the staggered magnetization at $r < 1/\Delta$
decays only as the first power of distance, see Eq. (\ref{eq:phi_RG}).
This is why a tiny concentration of impurities can significantly
influence the critical behaviour of the system.

Finally, we have analyzed the semiclassical limit  of a very large impurity
spin, $S \gg 1$. In this limit the quantum spin-flip transitions become
negligible and the spin impurity problem  is reduced to an exactly solvable 
textbook example. \cite{Mahan}

\acknowledgements
We gratefully acknowledge A. Milstein and H. Scammell for useful discussions. This research was supported by Australian Research Council (Grant No. DP110102123).

\end{document}